\documentclass[final]{agujournal}
\draftfalse

\newcommand{\degree}{^{\circ}}

\journalname{JGR-Planets}

\begin{document}

%
%


\title{Expected seismicity and the seismic noise environment of Europa}

%
%




\authors{Mark P. Panning\affil{1,2}, Simon C. St\"{a}hler\affil{3}\thanks{now at Institute of Geophysics, ETH Z\"{u}rich, Z\"{u}rich, Switzerland}, Hsin-Hua Huang\affil{4}\thanks{now at Institute of Earth Sciences, Academica Sinica, Taipei, Taiwan}, Steven D. Vance\affil{2}, Sharon Kedar\affil{2}, Victor C. Tsai\affil{4}, William T. Pike\affil{5}, Ralph D. Lorenz\affil{6}}


\affiliation{1}{Dept. of Geological Sciences, Univ. of Florida, Gainesville, Florida, USA}
\affiliation{2}{Jet Propulsion Laboratory, California Institute of Technology, Pasadena, California, USA}
\affiliation{3}{Dept. of Earth and Environmental Sciences, Ludwig-Maximillians-Universit\"{a}t, Munich, Germany}
\affiliation{4}{Seismological Laboratory, California Institute of Technology, Pasadena, California, USA}
\affiliation{5}{Department of Electrical and Electronic Engineering, Imperial College, London, United Kingdom}
\affiliation{6}{Space Exploration Sector, Johns Hopkins Applied Physics Laboratory, Laurel, Maryland, USA}




\correspondingauthor{M. P. Panning}{mpanning@ufl.edu}




\begin{keypoints}
\item Seismic activity level and ambient seismic noise due to tidal cracking in the ice on Europa is estimated 
\item Activity and noise is modeled with a Gutenberg-Richter relationship using numerical wave propagation
\item Autocorrelation of seismic noise may potentially be used to determine structure in the absence of large events.
\end{keypoints}

%
%


\begin{abstract}
Seismic data will be a vital geophysical constraint on internal structure of Europa if we land instruments on the surface.   Quantifying expected seismic activity on Europa both in terms of large, recognizable signals and ambient background noise is important for understanding dynamics of the moon, as well as interpretation of potential future data. Seismic energy sources will likely include cracking in the ice shell and turbulent motion in the oceans.  We define a range of models of seismic activity in Europa's ice shell by assuming each model follows a Gutenberg-Richter relationship with varying parameters.  A range of cumulative seismic moment release between $10^{16}$ and $10^{18}$ Nm/yr is defined by scaling tidal dissipation energy to tectonic events on the Earth's moon.  Random catalogs are generated and used to create synthetic continuous noise records through numerical wave propagation in thermodynamically self-consistent models of the interior structure of Europa.  Spectral characteristics of the noise are calculated by determining probabilistic power spectral densities of the synthetic records.  While the range of seismicity models predicts noise levels that vary by 80 dB, we show that most noise estimates are below the self-noise floor of high-frequency geophones, but may be recorded by more sensitive instruments.  The largest expected signals exceed background noise by $\sim$50 dB.  Noise records may allow for constraints on interior structure through autocorrelation.  Models of seismic noise generated by pressure variations at the base of the ice shell due to turbulent motions in the subsurface ocean may also generate observable seismic noise.
\end{abstract}

%
%

%


%
%
%
%


\section{Introduction}
Europa is a fascinating target which is likely to be a focus of future planetary science missions.  Observations from the Voyager and Galileo missions \citep{Kohlhase+1977,Russell2012} and Earth-based observations reveal a young, fractured icy surface \citep{Zahnle+2003} with magnetic signals that indicate a global subsurface ocean \citep{Kivelson+2000}.  The presence of liquid water makes the moon a prime target for astrobiological investigations, leading to multiple future missions in the planning stages, including the Europa Clipper from NASA \citep{Phillips+2014}, the JUICE mission from ESA \citep{Grasset+2013}, and a proposed Europa lander mission \citep{Hand+2017}.

While coarse information about subsurface structure of planetary bodies can be gleaned from gravity and magnetic observations from orbital and Earth-based observations, such models are generally quite non-unique.  Other geophysical observations will be required in order to determine more precise details, such as thickness of the ice shell, depth to the ocean bottom, and any other details of structure beneath the ocean.  Orbital ice penetrating radar measurements have long been proposed to look for the depth to Europa's ocean \citep[e.g.][]{Chyba+1998}.  Scattering in an impact gardened or tidally fractured regolith may cause this to be a problematic observation \citep{Eluskiewicz2004}, although more recent studies suggest the scattering may not be so problematic \citep{Aglyamov+2017}.  If a lander is sent to Europa \citep{Hand+2017}, a seismometer would provide an important complement to radar observations.  In fact, seismology has been the primary geophysical technique constraining the detailed structure of the Earth's interior.  Several previous studies have identified seismic signals ranging across a broad frequency band with the potential to provide important subsurface structural constraints on Europa \citep{Kovach+2001,Lee+2003,Panning+2006b}, while seismic investigations also present the opportunity to observe other signals of activity in Europa's ice shell and ocean that could have relevance to astrobiological investigations \citep{Vance+2017a}.

Understanding the seismic energy budget for Europa is both an important constraint on the ongoing dynamic processes of the planetary body, as well as a critical constraint on both ``signal'' (large, temporally and spatially isolated sources) and ``noise'' (nearly continuous and ongoing seismic excitation) for any potential seismic recording.  In order to use seismology to increase our understanding of Europa, it is important to attempt to quantify instrument requirements as well as possible.  Obviously, an instrument needs to be sensitive enough to record the desired signals, whether these be body waves reflecting from the base of the ice shell and ocean \citep[e.g.][]{Lee+2003} or other longer-period signals \citep{Kovach+2001,Panning+2006b}, but it is also important to estimate the amplitude of the ambient noise.  This noise estimate informs mission planners of likely signal-to-noise ratios and sets an important baseline for instrument sensitivity.  Designing an instrument that is sensitive to signals orders of magnitude below the ambient noise floor is inefficient, but being able to record the noise floor can provide important science return on its own.  

On Earth, the majority of the energy of the ambient noise field originates in the oceans, and analyzing it in the ``microseismic band'' (at periods roughly between 5 and 20 seconds) provides important constraints on wave interactions relatively near the shore \citep[e.g.][]{Longuet-Higgins1950}, while signals due to infragravity waves excited by ocean storms cause continuous oscillations of the whole Earth at much longer periods \citep[e.g.][]{Rhie+2004}.  Constraining the amplitudes and frequency characteristics of the noise therefore can give us important constraints on active processes occurring on a planetary scale.  Following work looking at correlation of noise in event codas producing signals that approximate the seismic response between two stations \citep{Campillo+2003}, noise records on Earth have also been extensively used in the last decade and a half to generate useful signals to constrain internal structure.

On Europa, we do not yet have any direct constraints on the seismic signal and noise characteristics, but it is reasonable to hypothesize that such noise will be primarily generated by a combination of widespread ice-tectonic events (i.e. tidal cracking) within the ice shell and motions of the subsurface ocean.  The energy source for both types of noise would be tidal deformation due to the slightly elliptical orbit of Europa around Jupiter. There is observational evidence for Hubble Space Telescope imaging of transient plumes of water vapor \citep{Roth+2014} which suggests that opening and closing of cracks is ongoing today, and is controlled by tides \citep{Rhoden+2015}. Further observations recently confirmed the plumes \citep{Sparks+2017} and suggested they may be correlated with a thermal anomaly associated with tidal friction and/or access to the internal ocean. 

In this study, we propose to produce quantitative estimates of the seismic signal and noise due to widespread small ice-tectonic events, which is a problem amenable to careful quantification with a small number of assumed properties of Europa using available seismic modeling tools.  With this noise estimate, we demonstrate some potential for using autocorrelation of noise recording to look for ice shell and ocean geometry. Finally, we also look at an initial estimate of possible noise due to one model of turbulent flow in Europa's ocean \citep{Soderlund+2014}, although further work on this noise source will be important.

\section{Modeling tidal cracking events with a Gutenberg-Richter relationship}
\label{catalogs}
The slightly elliptical orbit of Europa around Jupiter causes tidal stresses that vary diurnally with amplitudes of $\sim$100 kPa \citep[e.g.][]{Hoppa+1999}.  This stress has been modeled as sufficient to induce fracturing extending 10's of meters into the ice shell \citep{Lee+2003}.  In addition, some surface observations are used to argue for non-synchronous rotation of Europa \citep[e.g.][ for further details]{Bills+2009}, and such motions could cause larger stress values, possibly exceeding 1 MPa \citep[e.g.][]{Hurford+2007,Beuthe2015}.  Such larger stresses could cause cracking extending several kilometers into the ice shell \citep{Lee+2005}, which would generate significant seismic energy release.

Rather than attempting to model the distribution of ice-cracking events through detailed stress modeling, which depends strongly on many assumptions such as ice rheology, porosity, and brine content, we choose instead to use a simple statistical model to determine seismic energy release due to ongoing cracking events in the Europan ice shell.  Following work by \citet{Golombek+1992} for modeling seismicity of Mars, we choose to assume seismicity in Europa's ice shell will follow a Gutenberg-Richter relationship \citep{Gutenberg+1944}, which is typically written as a log-linear relationship between the number of events observed greater than or equal to a particular earthquake magnitude,  
\begin{equation}
\log{N(M_W)} = a - b M_W,
\label{GRdefine}
\end{equation}
where $N(M_W)$ is the number of events greater than or equal to moment magnitude $M_W$, and $a$ and $b$ are empirically defined parameters which are fit to a particular seismicity catalog.  Moment magnitude can be related to the seismic moment, $M_0$, of a particular seismic event by the definition of $M_W$ \citep{Kanamori1977},
\begin{equation}
\log{M_0} = 1.5 M_W + 9.1,
\label{MWdefine}
\end{equation}
and so \citet{Golombek+1992} chose to write the Gutenberg-Richter relationship in an equivalent form as
\begin{equation}
N(M_0) = AM_0^{-B},
\label{G92-1}
\end{equation}
where $A$ and $B$ are empirical parameters, which can be related to $a$ and $b$ by substitution from equation~\ref{MWdefine} into equation~\ref{GRdefine} to obtain $a=\log{A} - 9.1B$ and $b=1.5B$.

If we assume seismicity on Europa will follow such a relationship (as is true for observed catalogs on the Earth and the Moon), expected numbers of events of any size can be calculated simply by specifying physical parameters that constrain the parameters $A$ and $B$.  \citet{Golombek+1992} showed that these can be uniquely determined by specifying the cumulative seismic moment released per year, $\Sigma M_0$, the maximum event size, $M_0^\star$, and a value for $b$, which specifies the slope in the Gutenberg-Richter relationship, where the cumulative seismic moment and maximum event size can be related to $A$ and $B$ by the relationship
\begin{equation}
\Sigma M_0 = \frac{AB}{1-B}\left(M_0^\star\right)^{1-B}.
\label{G92-2}
\end{equation}
While some variation in $b$ values is observed in different seismic catalogs, all Earth catalogs generally have values that vary between $\sim$0.7 and $\sim$1.3 \citep{Frohlich+1993}, and so we choose to simply assume $b=1$ for most catalogs we develop in this study, leaving us only to define the cumulative moment release and maximum event size.  For lunar records, though, the $b$ value may differ from this narrow range. \citet{Lammlein+1974} suggested a very high $b$ value of 1.78 for waveform identified tectonic events based on the logarithm of observed amplitudes rather than earthquake magnitude, while \citet{Nakamura1977} observed a very low $b$ value of 0.5 for a catalog of the largest distant events, which he categorized as High Frequency Teleseismic (HFT) events. Obviously, these still bracket a value of 1, but represent very different end members.  Additionally, it is possible that material properties and rheology of ice may lead to different $b$ values for ice tectonics as compared to tectonics in silicate materials.   Terrestrial studies of icequakes in various environments cover a wide range of apparent $b$ values \citep[see review by][]{Podolskiy+2016}.  Most studies, however, either show a value of near 1 or clustered at higher values approaching 2, which implies a much greater number of small earthquakes for a given seismic activity rate.  For non-fracture related ice seismic sources, like calving of glaciers, events may not follow a Gutenberg-Richter distribution at all, but instead show a characteristic event size \citep[e.g.][]{Veitch+2012}.  For this study, we choose to primarily focus on a $b$ value of 1, but explore implications of varying $b$ value in section~\ref{bvalue}.

We consider a range of options for the cumulative moment release estimate.  A reasonable starting point for such an estimate would be to scale it to our only available planetary catalog aside from the Earth.  Based on Apollo data, lunar seismicity is described by a cumulative seismic moment release of approximately $10^{15}$ Nm/yr \citep{Oberst1987}.  While there are multiple models for the driving energy of lunar seismicity, tidal periodicities in the occurrence rate \citep[e.g.][]{Lammlein+1974} suggest an important role for tidal dissipation energy in driving quakes.  The dissipated tidal energy in the moon is estimated at 1.36 GW \citep{Williams+2001}, while the dissipated energy in the ice shell of Europa has been estimated from 630 GW up to a few thousand GW \citep{Tobie+2003,Hussmann+2004,Vance+2007}, which is larger by approximately 3 orders of magnitude.  Based on this, it is reasonable to expect cumulative moment release on Europa to significantly exceed that of the Earth's moon.  However, it is likely an oversimplification to simply assume a linear scaling.  For example, while brittle fracture leading to seismic energy release is likely in the upper portion of the icy shell \citep[e.g.][]{Lee+2003,Lee+2005}, the ice will likely be ductile at greater depths and higher temperatures, which means that energy dissipated at these depths is unlikely to produce seismic moment release.  Based on spacings of geologic features, the brittle-ductile transition on Europa has been placed at a depth of approximately 2 km \citep[e.g.][]{Pappalardo+1999}, meaning that the brittle portion of the shell would make up anywhere from several percent of the total volume for a thick shell to several tens of percent of the volume of a thin shell, which would suggest a corresponding reduction to simple linear scaling from the Earth's moon.  Based on these considerations, we make an initial approximation that activity levels will be 1 to 3 orders of magnitude above that of the moon, leading to a range of $10^{16}$ to $10^{18}$ Nm/yr.  Additionally, there may be other energy sources for lunar seismicity which complicate a simple scaling relationship.  For example, the HFT events, which are some of the largest recorded by the Apollo mission, do not appear to be linked to the tidal cycle.  As a matter of fact, the cause of these events is very uncertain, and has even been proposed to be linked to encounters with high velocity nuggets of strange quark matter \citep{Frohlich+2006}.  To be conservative, we are treating the simple scaling result as an upper bound, but also considering rates up to 2 orders of magnitude lower.

 \begin{table}
 \caption{Model parameters for seismicity models}
 \centering
 \begin{tabular}{l c c}
 \hline
  Model & $\Sigma M_0$ (Nm) & $M_0^\star$ (Nm) \\
 \hline
  A & $10^{16}$ & $10^{19.5}$ \\
  B & $10^{16}$ & $10^{18}$ \\
  C & $10^{18}$ & $10^{19.5}$ \\
  D & $10^{18}$ & $10^{18}$ \\
  Preferred & $10^{17}$ & $10^{18.5}$ \\
   \hline
 \end{tabular}
 \label{modeltable}
 \end{table}

\citet{Nimmo+2006} argue based on observed surface faulting in regions with Galileo data of sufficient resolution that observed faulting corresponds to a seismic moment magnitude of $M_W$ 5.3, assuming a low shear modulus of the ice due to regolith development.  If the shear modulus of the ice is closer to that of unfractured ice, this increases to a magnitude of $\sim M_W$ 6 \citep{Panning+2006b}.  Given that only a limited portion of the surface was investigated, this suggests the maximum event size should be at least $M_W$ 6.  With a $b$ value of 1, there are an order of magnitude fewer events for each unit increase in magnitude, however a unit increase in magnitude corresponds to an increase in energy by a factor of $\sim$30.  This means the largest events dominate the cumulative moment release, and so a larger maximum event size implies relatively fewer of the frequent small events most likely to be observed by a short duration surface landed experiment.  This seems like a counter-intuitive result, as there is a general expectation that the maximum event size in a particular catalog and the cumulative moment release should be correlated.  Likely, such a correlation is to be expected in most settings; however, there is no explicit physical relationship defining how such a correlation should be defined.  Even if we could define such a relationship, we would expect some variance between catalogs in how closely correlated the maximum event size would be with the cumulative moment release.  Therefore, we choose to consider a range of maximum event size between $M_W$ 6 ($\sim 10^{18}$ Nm) and a more conservative estimate of $10^{19.5}$ Nm such as that used by \citet{Golombek+1992} for Mars based on the maximum observed intraplate oceanic earthquakes on Earth.

\begin{figure}
\includegraphics[width=15pc]{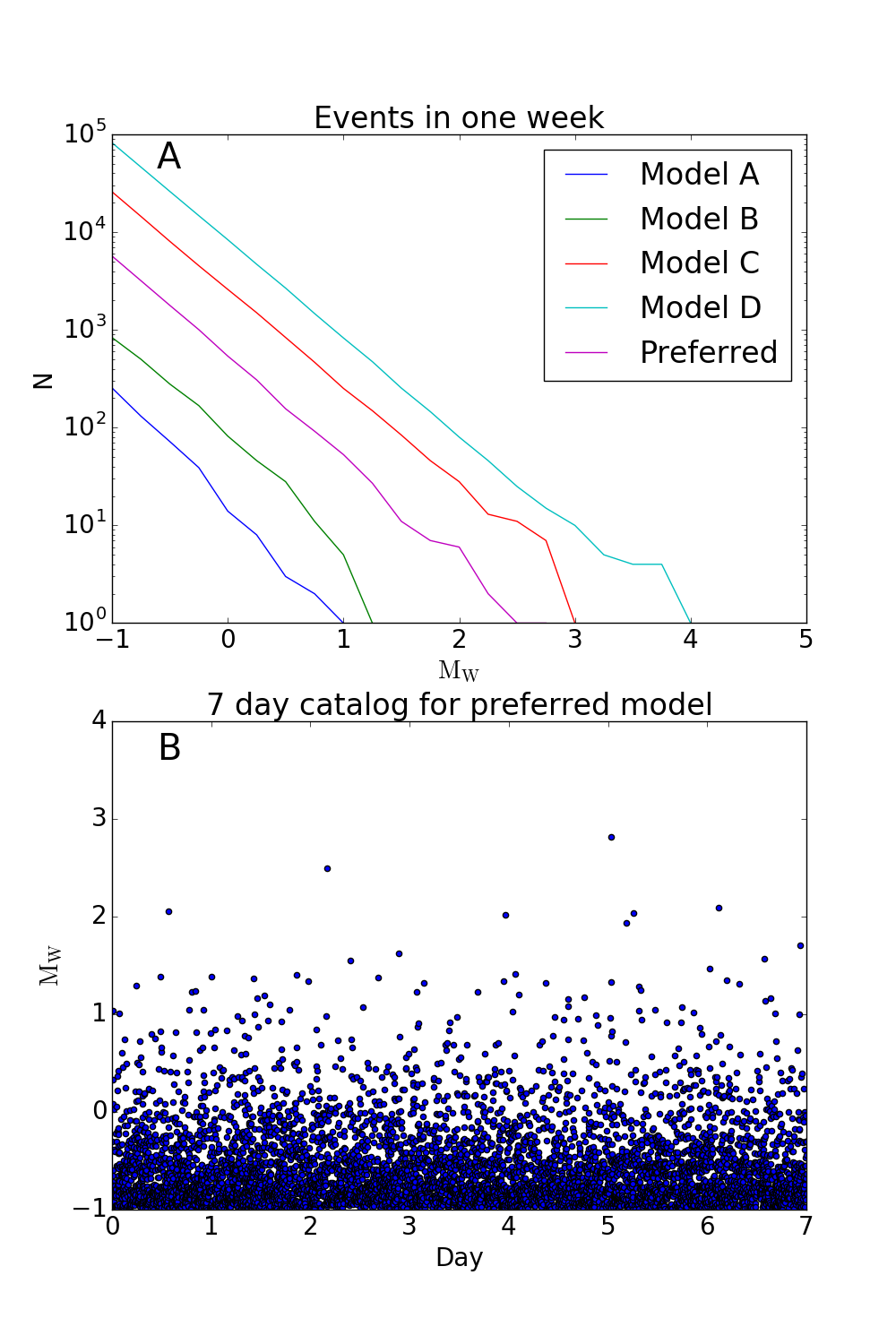}
\caption{(A) Plotted Gutenberg-Richter relationships for random one-week realizations of 5 different seismicity models described in table~\ref{modeltable}. (B) Plot of one week of randomly realized events for the ``preferred'' model.}
\label{catalogfig}
\end{figure}

Given this range of estimates for cumulative moment release, we choose to define 5 candidate models of Gutenberg-Richter parameters to describe the activity of ice-tectonic events in Europa's ice shell (fig.~\ref{catalogfig}).  We define 4 end member models with either low ($10^{16}$ Nm/yr; models A and B) or high ($10^{18}$ Nm/yr; models C and D) cumulative moment release, and either a large ($10^{19.5}$ Nm; models A and C) or small ($10^{18}$ Nm; models B and D) maximum event size (table~\ref{modeltable}).  Finally we defined a ``preferred'' model with parameters between the end members ($\Sigma M_0 = 10^{17}$ Nm and $M_0^{\star}=10^{18.5}$ Nm).  Figure~\ref{catalogfig}A displays the statistics of random 1-week realizations of these catalog parameters.  Theoretical Gutenberg-Richter relationships would make straight lines on these plots, but a random realization causes some variation around these straight lines, particularly near the small number of large events.  Each catalog is realized by calculating a probability of occurrence of each event size per second, and then creating a catalog by comparing these probabilities to random numbers generated for a desired length of time (fig.~\ref{catalogfig}B).  We can then use these catalogs to generate synthetic long-duration seismic records, provided we also randomly assign location and source mechanism characteristics to each event, as discussed in section~\ref{noise_estimates}.

\section{Europa structure models}
Accurately modeling potential Europa noise signals relies also on correctly characterizing the seismic wave propagation from modeled noise sources.  For this, we need realistic structure models that detail elastic properties, density, and anelastic attenuation structure.  \citet{Vance+2017} have produced a tool for building models of icy ocean worlds that are thermodynamically self-consistent, and include up-to-date thermodynamic properties for ices, saline oceans, as well as the rocky interior and iron core of Europa.  Radial structures are computed as per \citet{Vance+2014}, with self-consistent ice and ocean thermodynamics, using boundary conditions of surface and ice-ocean interface temperature. Thermodynamics for rock have been added as per \citet{Cammarano+2006}, with updates to account for rock porosity, mineral hydration, and the presence of Na-bearing minerals. All models are designed to match the observations of a bulk density of 2989$\pm$46~kg~m$^{-3}$ and normalized moment of inertia of 0.346$\pm$0.005 \citep{Schubert+2004}.  All modeling tools are freely available via GitHub (http://github.com/vancesteven/PlanetProfile).

\begin{figure}
\includegraphics[width=20pc]{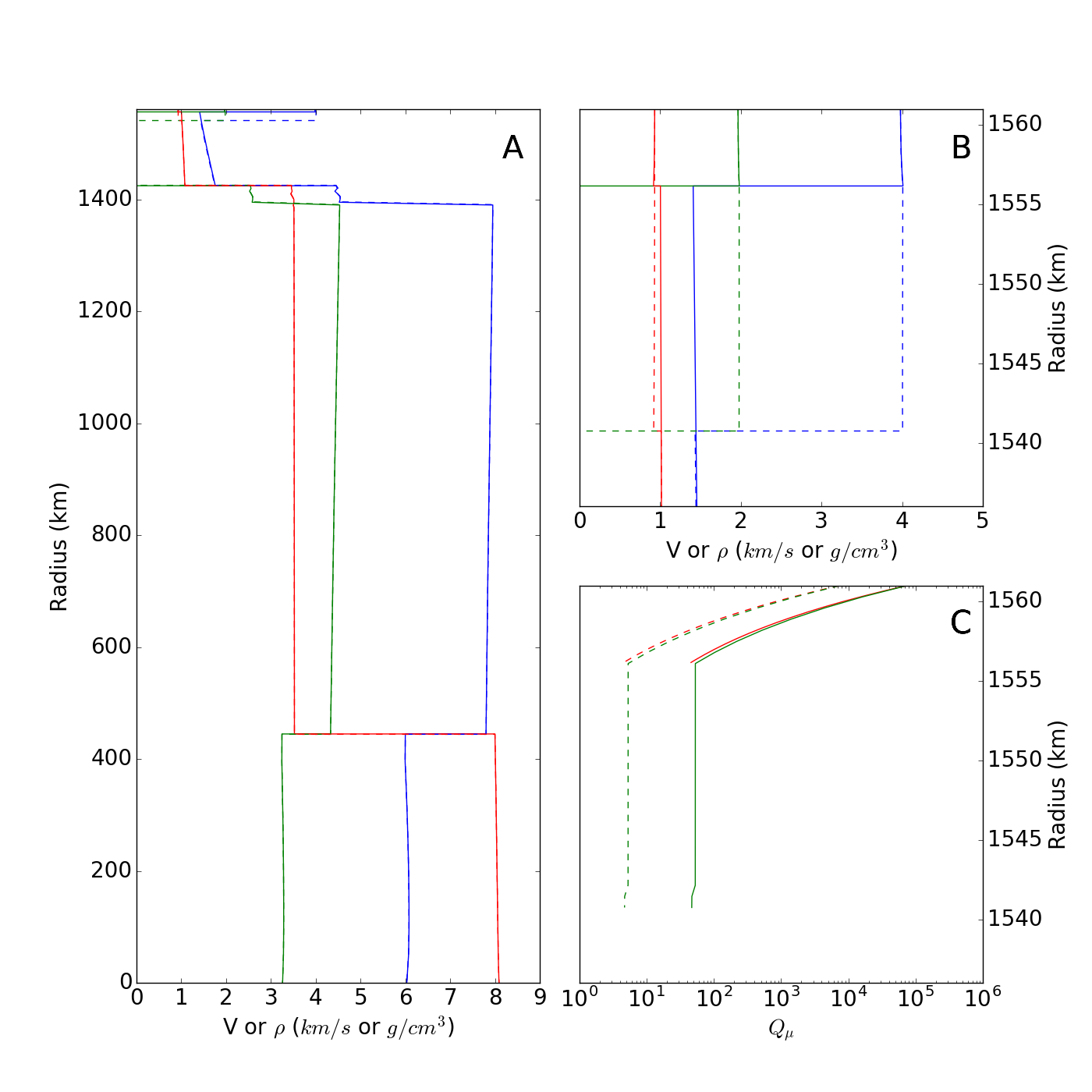}
\caption{(A) $V_P$ (blue), $V_S$ (green), and density (blue) for the 5 km thick ice shell model (solid) and 20 km thick ice shell model (dashed).  (B) Same as (A), but zoomed in to show ice shell and upper ocean structure.  (C) $Q_{\mu}$ values for the ice shells for the 5 km thick ice shell (red) and 20 km thick shell (green) for the high (solid) and low $Q$ (dashed) models.}
\label{modelfig}
\end{figure}

Temperature profiles can be tuned to produce different ice shell thicknesses.  It is also possible to consider a range of internal compositions and temperature profiles for the interior below the ocean, but we only explore differences in ice shell thickness in this study (fig.~\ref{modelfig}).  Previous modeling showed that observable seismic signals for sources located within the ice shell are dominated by the structure of the ice shell and have little sensitivity to structure below the ocean \citep{Panning+2006b}.  For this study, we consider two different ice shell thicknesses, 5 and 20 km.  Ice shell thickness has a very strong influence on the character of the surface waves, which grade from relatively non-dispersive Rayleigh waves to flexural waves at a characteristic frequency that depends on the thickness.  There are also guided SV waves in the ice shell \citep[Crary waves,][]{Crary1954} that have characteristic frequency content that depends strongly on ice shell thickness.  While ice shells thicker than 20 km are possible, these two values roughly bracket a reasonable range to constrain how overall noise characteristics may vary with thickness.

Amplitudes also depend on the attenuation structure of the model.  For initial estimation, we followed the approach of \citet{Cammarano+2006} to obtain temperature and frequency dependent estimates of shear quality factor, $Q_{\mu}$ with the expression
\begin{eqnarray}
\frac{Q_{\mu}}{\omega^\gamma} &=& B_a\exp\bigg(\frac{\gamma H(P)}{RT}\bigg) \label{anelastic}\\
H(P) &=& g_a T_m,
\end{eqnarray}
in which $B_a=0.56$ is a normalization factor, $\omega$ is the seismic frequency, exponent $\gamma=0.2$ is the frequency dependence of attenuation, and $R$ is the ideal gas constant. $H$, the activation enthalpy, scales with the melting temperature $T_m$ and with the anisotropy coefficient $g_a$, and the values of $g_a$ chosen for various ices are described in \citet{Vance+2017}.  The bulk quality factor, $Q_{\kappa}$, is neglected.  This relationship predicts very high $Q$ values, and therefore very little attenuation, within the ice shell (fig.~\ref{modelfig}C).  However, attenuation in ice at very low temperatures is not very well constrained. Although studies of glacier ice suggest near-surface layers can have very low $Q$ \citep{Gusmeroli+2010}, high $Q$ values are reached at low temperatures.  Fractured ice may also be more attenuating than simple melting temperature scaled solid ice estimates, and partial but incomplete saturation with fluids can lower $Q$ even further \citep{Peters+2012}.  For these reasons, we choose to use two different $Q$ structures: one predicted by equation~\ref{anelastic}, and one with $Q$ arbitrarily reduced by a factor of 10.  Combining the 2 different ice shell thicknesses and 2 different $Q$ structures explored, we have 4 structure models to explore in this study.  Combined with the 5 seismicity models discussed in section~\ref{catalogs}, we have 20 different noise simulations to create.

\section{Noise estimates}
\label{noise_estimates}
\begin{figure}
\includegraphics[width=20pc]{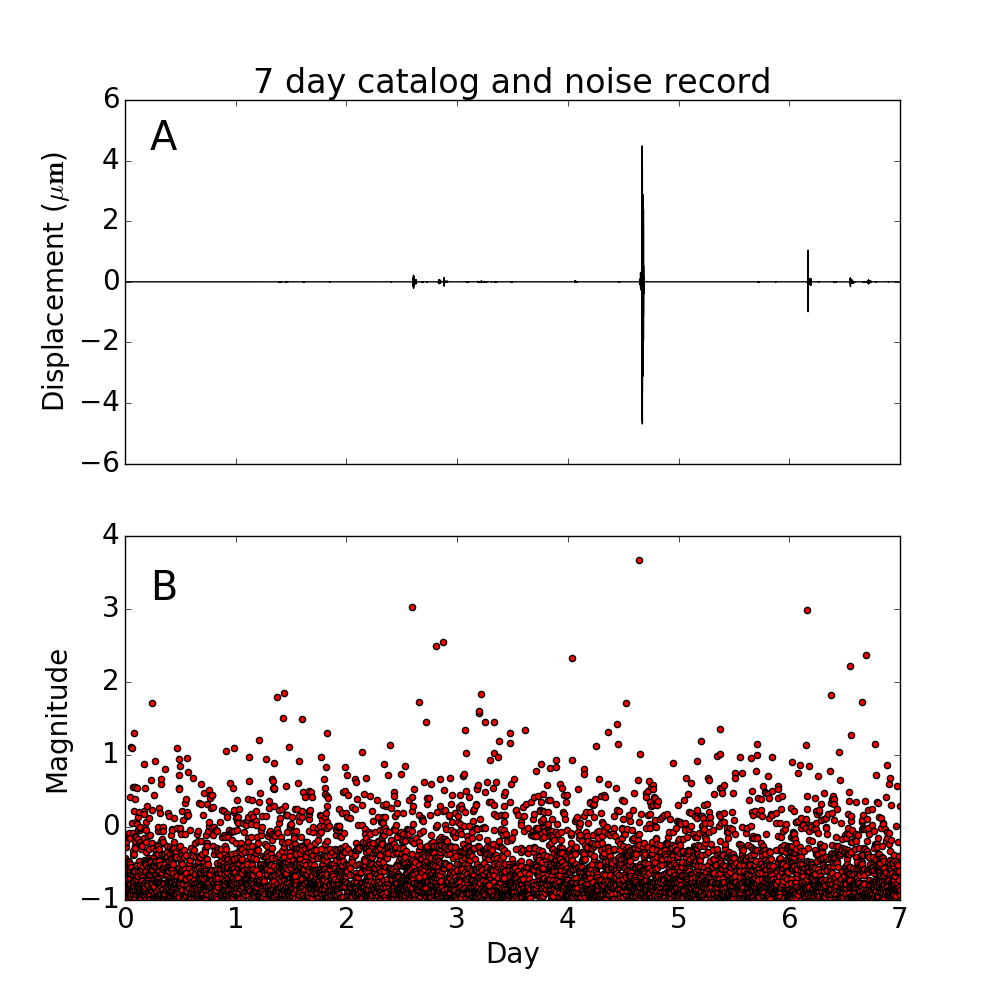}
\caption{Sample noise record (A) plotted for a 1 week realization of the ``preferred'' seismicity model (B) in the 5 km ice shell model with high $Q$.}
\label{noisecatalogfig}
\end{figure}

With the catalogs based on assumed Gutenberg-Richter relationships and the thermodynamically self-consistent models in place, we have the ingredients to create synthetic noise records representing the expected background noise due to cracking events in the ice.  In order for these to be useful, we need to be able to rapidly calculate waveforms for arbitrary source locations up to frequencies near 1 Hz for thousands to tens of thousands of events.  Fortunately, the python-based Instaseis program \citep{vanDriel+2015} is well-suited for problems of this type.  Instaseis makes use of full waveform databases computed with the axisymmetric spectral element code AxiSEM \citep{NissenMeyer+2014}.  These databases are computed for 2 sources (one vertical force and one horizontal) located at the surface at the north pole in a 1D spherically symmetric planetary model, which can then be rapidly interpolated to arbitrary source and receiver geometries using the principle of reciprocity.  This allows for rapid seismogram calculation, from milliseconds to seconds on a desktop processor depending on length and frequency content of the waveform database, although the initial waveform database is a larger computational investment.  AxiSEM is readily able to handle arbitrary planetary models, as demonstrated by the exploration of ocean world seismology by \citet{Stahler+2017}.

\begin{figure}
\includegraphics[width=15pc]{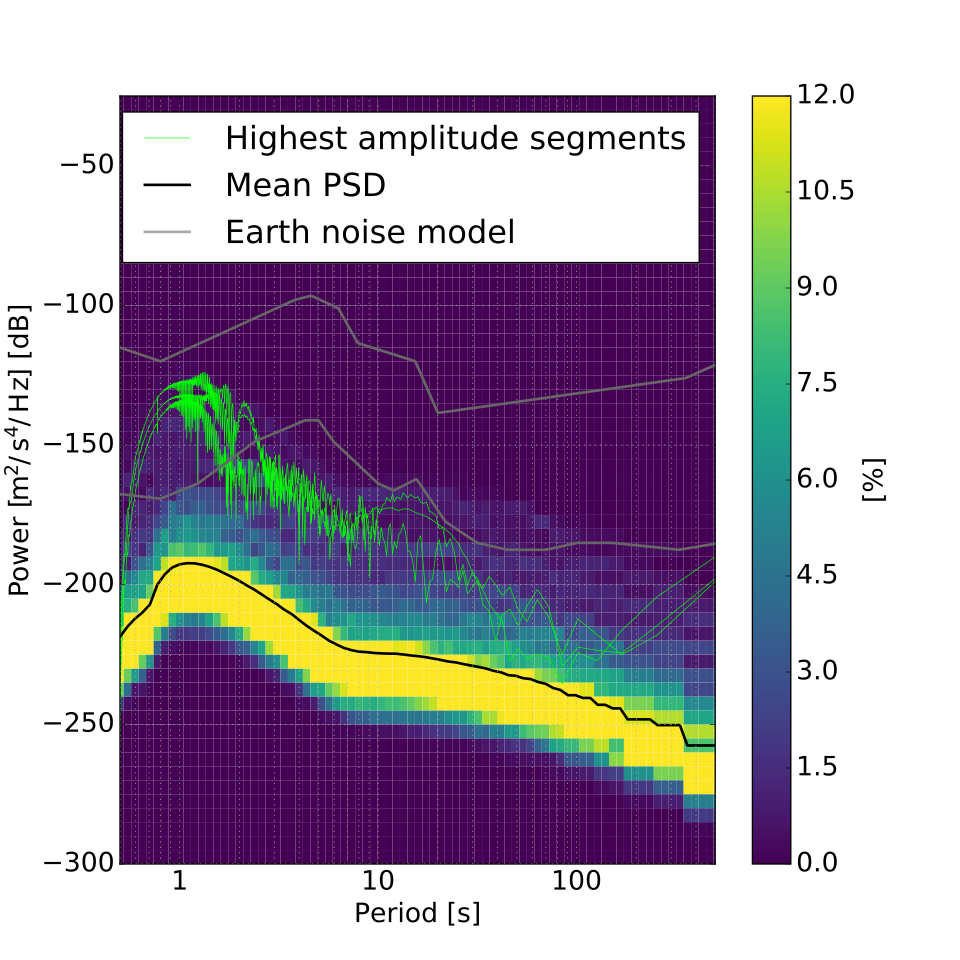}
\caption{An example probabilisitic power spectral density for the 5 km thick ice shell high Q model with the preferred seismicity model.  Background colors are the probability density function of ground acceleration power using 584 partially overlapping 1 hour segments covering 2 weeks of simulated noise records.  The solid black line represents the mean PSD, while the thin green lines represent the PSDs for the highest three 1 hour segments.  Note that the peak amplitudes are 40--50 dB above the mean background noise.  Grey lines represent the low and high noise models for background noise observed at Earth stations \citep{Peterson1993}.}
\label{ppsdfig}
\end{figure}

The catalogs of section~\ref{catalogs}, which initially were only calculated to give a time-series of quake occurrence, need to be completed by specifying all the relevant source characteristics.  For this exercise, we assumed a polar location for our station and a homogenous distribution of epicenters on the surface of the planet.  Depth was randomly assigned between the surface and 2 km depth, a commonly assumed depth of the brittle-ductile transition \citep{Pappalardo+1999}.  Strike, rake, and dip were also randomly assigned.  Clearly, the real seismicity on Europa will likely be influenced by the tidal stress pattern, causing variations in seismicity rate in both time and space, but a homogenous distribution was chosen as an initial baseline estimation of the kind of seismic activity we could expect for a landed seismometer on the surface of Europa.  Once these source characteristics are defined for the catalog, we generate noise records using Instaseis with all events calculated using AxiSEM databases with 1 hour databases calculated to a dominant frequency of 1 Hz. 
Figure~\ref{noisecatalogfig} displays a typical seismic trace realized for a 1 week catalog based on the ``preferred'' seismicity model.  Note that the record is dominated by a handful of larger events, which is typical for any record calculated with a Gutenberg-Richter relationship with a $b$ value of 1.  Between these larger events, however, a background level of seismic energy develops from the large numbers of smaller events (e.g. figures~\ref{catalogfig}B and \ref{noisecatalogfig}B).  As another way of presenting such simulated records, we have created sound files by speeding up portions of the records by a factor of 500 using publicly available Matlab tools \citep{Kilb+2012}, and included these in supporting information.

In order to determine reasonable spectral characteristics of the average power of the ambient noise from this record, we need to be careful to not simply estimate power spectral density of the whole record.  This will be dominated by the sporadic large events, and not represent the power in the noise between these events, which is what the sensor will record the majority of the time.  To account for this, we use a probabilistic approach to determining the power spectral density (PSD) that is commonly used when assessing noise levels recorded by seismic stations on Earth \citep{McNamara+2004b}.  This is implemented by the PPSD tool in the signal processing toolkit of ObsPy \citep{Krischer+2015}, which is a seismic package for Python.  In this approach, the record is divided into a series of overlapping 1 hour segments (the value typically used in evaluating noise characteristics of Earth stations), and a PSD is determined for each segment.  These are then stacked in order to obtain a probability density function of the noise (fig.~\ref{ppsdfig}).  We note that in a Gutenberg-Richter relationship, the choice of window length in PSD estimates does have an impact on estimates of noise level.  We choose 1 hour windows to be consistent with standard evalutations, and to insure we cover the relevant seismic frequency band, but we evaluated a shorter window as well.  If we use a window of 600 seconds instead, all estimates of mean background noise become lower by several dB.  This means that even on the scale of an hour, there is variation in the signal level, and we are not seeing something that can be characterized as a completely stationary, random process.

\section{Results}
\label{resultssec}
\begin{figure}
\includegraphics[width=20pc]{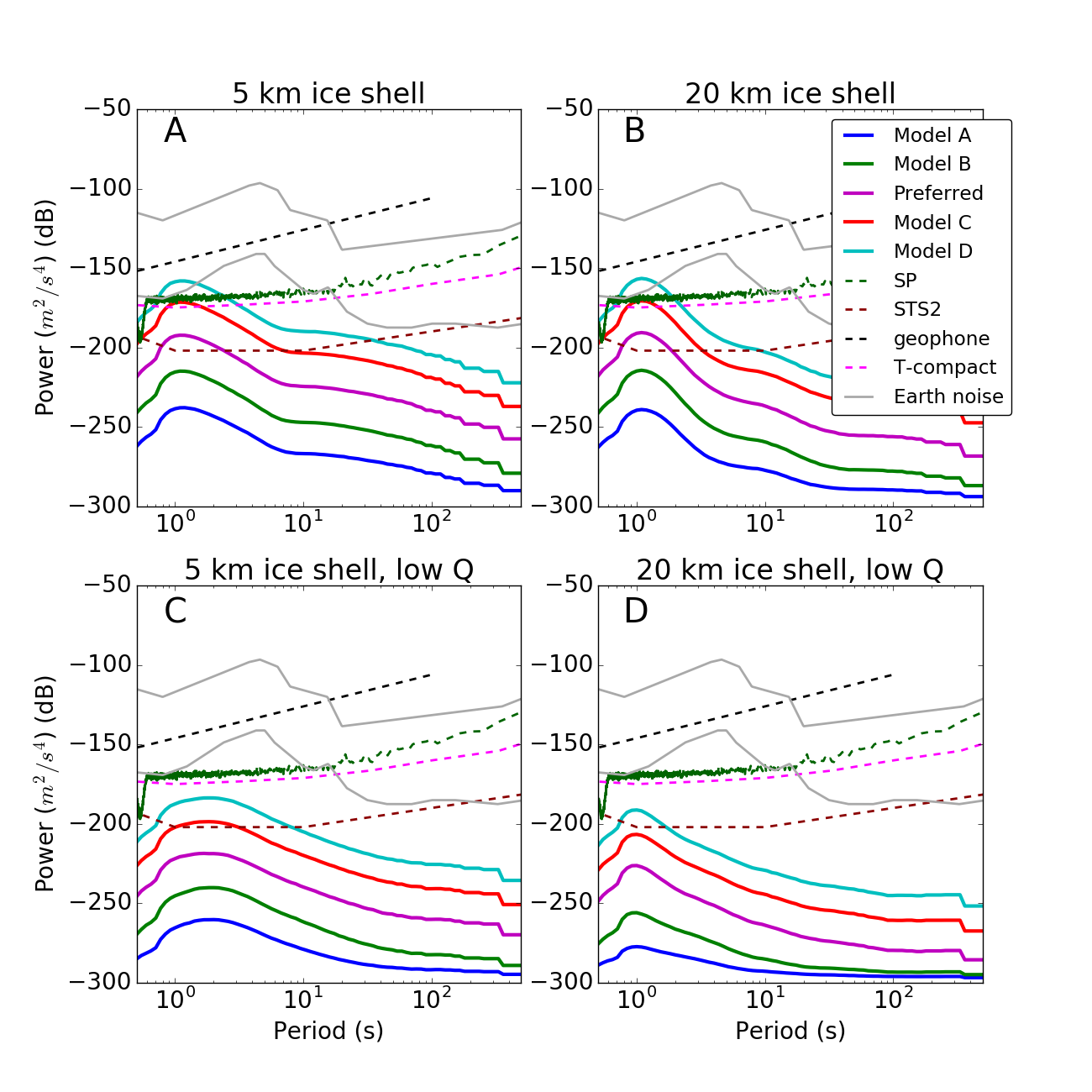}
\caption{Summary noise figures plotting mean PSD estimates for noise models in 5 km thick (A and C) and 20 km thick ice shell models (B and D) assuming high $Q$ (A and B) or low $Q$ in the ice shell (C and D).  For each model, the mean PSD estimate is plotted for the five seismicity models as solid lines.  Low and high noise models for the Earth are plotted in grey \citep{Peterson1993}.  For comparison, self-noise curves for broadband STS2 and Trillium Compact Earth instruments \citep{Ringler+2010}, the SP instrument which will be launched with the InSight mission \citep{Pike+2016} and a typical high frequency geophone \citep{Rodgers1994}.}
\label{noisesummary}
\end{figure}

For each combination of our 4 structural models and 5 seismicity models, we calculated a probabilistic PSD estimate of the ground acceleration noise power analogous to fig.~\ref{ppsdfig}.  While we calculated full 3 component noise series, we choose to focus in this study on the vertical component, which is frequently least affected by local site effect noise in terrestrial applications.  In order to facilitate plotting multiple estimates on a single axis, we instead plot only the mean value (i.e. the black line in fig.~\ref{ppsdfig}) for all seismicity models on the same axis for each structural model (fig.~\ref{noisesummary}).  To give context to these noise estimates, we also plot the low and high noise models for the Earth \citep{Peterson1993}, as well as self-noise models for several seismic instruments, ranging from an industry standard for high quality broadband instruments \citep[the STS2 with the dark red dashed line,][]{Ringler+2010} to a readily available low cost high frequency geophone \citep[dashed black line,][]{Rodgers1994}.  In between are the noise estimates for a Trillium Compact instrument \citep{Ringler+2010}, which is a common instrument used in Earth applications, and the SP instrument built for the InSight mission to Mars, due to launch in 2018 \citep{Pike+2016}.  These two instruments have a similar noise floor between the top-of-the-line broadband instruments and the high-frequency geophones.

For the overall noise level, both the 5 km and 20 km thick ice shell models produce similar amplitudes at the highest frequencies we explored near 1 Hz (fig.~\ref{noisesummary}).  The thinner ice shell models have higher amplitudes at lower frequencies, with a difference of 10--20 dB near periods of 10 s, depending on whether we are comparing the low or high $Q$ models.  This is consistent with the signal of large amplitude lower frequency flexural waves predicted for thinner ice shells \citep{Panning+2006b}.  As expected, the lower $Q$ models also predict lower amplitude noise models, with the factor of 10 difference in $Q$ here leading to approximately 20 dB lower power signal near 1 Hz in the low $Q$ models compared to the high $Q$ models of the same thickness.  The range of seismicity models explored here, however, are the biggest source of uncertainty.  The difference in signal power between the high seismicity model D and low seismicity model A leads to a $\sim$ 80 dB offset of our final noise power estimates.

Compared with the instrument noise curves, we can see that a relatively low sensitivity geophone is unlikely to record the ambient noise due to tidal cracking, regardless of the model chosen or the seismicity level.  More sensitive instruments like the InSight SP may be able to record this kind of ambient noise for higher overall seismicity levels, at least near 1 Hz, and a very sensitive instrument may be able to record over a broader frequency range out to 10 s period for the highest seismicity levels.

\begin{figure}
\includegraphics[width=25pc]{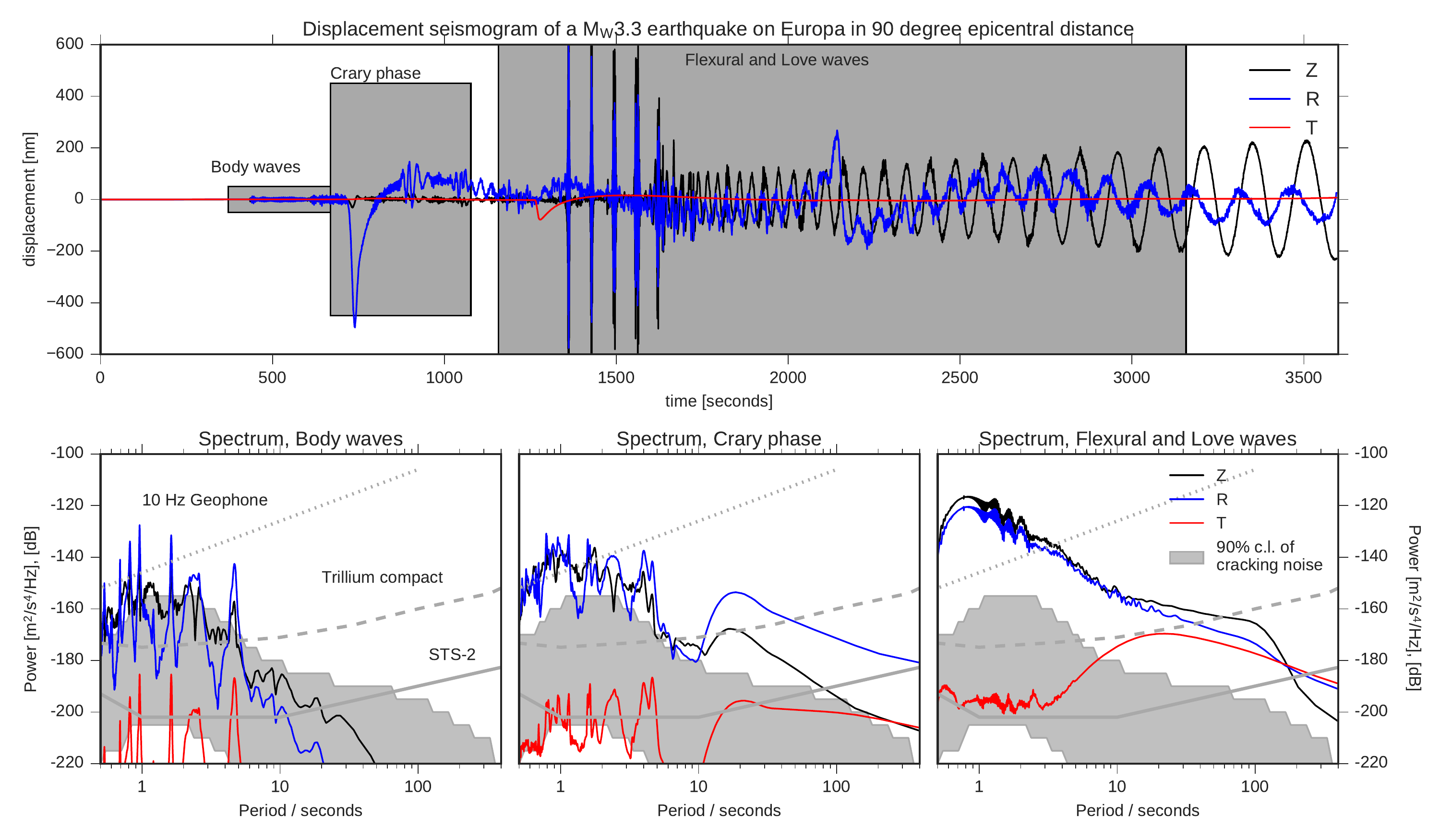}
\caption{Sample three component seismogram (vertical component black, horizontal component along the great circle between source and receiver in blue, and perpendicular to the path in red) for a $M_W$3.1 event at a depth of 2 km at a 90$\degree$ epicentral distance from the station in the 5 km ice shell high $Q$ model (top).  Approximate scattering in the ice shell is included (see section~\ref{scattering_sec} for details).  Bottom row shows spectra for the three grey windows for the three components calculated by averaging over 100 random focal mechanisms.  Grey field in lower panels represents the 90\% confidence limit from the PPSD calculated in model C.}
\label{phase_spectra}
\end{figure}

The peak recorded signals, though, representing the largest events during the span of 2 to 3 weeks, rise 30--50 dB above the mean noise level (fig.~\ref{ppsdfig}), and are thus likely recordable between 1 Hz and 10 s period with an instrument similar in quality to the SP instrument or Trillium compact.  This separation of 30-50 dB is actually a conservative estimate, as it is based only on 1 hour calculations for the PSD.  A window chosen to specifically highlight the event would return a slightly higher power estimate.  In fact, specific phases of interest like body waves which may record ice shell and ocean reflections \citep[e.g.][]{Lee+2003}, or the Crary phase which is sensitive to ice shell thickness \citep[e.g.][]{Vance+2017a}, or the surface waves, including flexural waves \citep[e.g.][]{Panning+2006b} rise above both instrument noise curves and background noise levels (fig.~\ref{phase_spectra}).  The traces shown in figure~\ref{phase_spectra} rise above the background noise even though they include an approximation of the effects of moderate scattering (see section~\ref{scattering_sec} for details about how this is implemented).  These are calculated for a moment magnitude ($M_W$) 3.1 event 90$\degree$ from the lander.  This is a reasonable estimate for the largest event recorded in a few weeks given our preferred seismicity model (fig.~\ref{catalogfig}).  Even these high amplitude arrivals, though, only just begin to approach the self-noise floor of the high frequency geophone.  This estimate, however, is based on a relatively distant event, which is reasonable if we assume a homogenous, random distribution of events with a landing site that is not chosen to maximize probability of recording an event.  In this case the vast majority of events recorded will take place between 45$\degree$ and 135$\degree$ from the lander based simply on surface area of a sphere.  If a landing site on Europa (or similarly for Enceladus or Titan) were chosen to be close to areas of observed activity (such as observed plumes or modeled maximal tidal stresses), it may be reasonable to expect larger events closer to the lander, and a less sensitive instrument may be sufficient.  Further modeling of expected activity would then be essential.

\section{Discussion}
\subsection{Autocorrelation of ambient noise}
While our initial results suggest ambient noise due to cracking events alone may be hard to reliably record, there will likely be other noise sources such as ocean noise (see section~\ref{oceannoise}).  Regardless of the source of noise, we may be able to use reliable recordings of background noise (i.e. background noise above the instrument self-noise) to extract useful information about structure, even in the absence of identifiable larger ice tectonic events.  \citet{Claerbout1968} suggested that  autocorrelation of ambient noise should produce the equivalent reflection response as if you had a source co-located with the receiver, and this approach has been applied numerous times since then using either earthquake coda \citep[e.g.][]{Wang+2015,Huang+2015} or ambient seismic noise \citep[e.g.][]{Tibuleac+2012,Kennett2015,Saygin+2017}.  \citet{Zhan+2013} explored the use of noise autocorrelation in a potentially analogous setting for a deployment of broadband seismometers at 24 different sites on the Amery Ice Shelf, Antarctica.  While for most sites, the incoherent noise (e.g. the mechanical and electrical noise of the station installation) exceeded the coherent noise, the authors were able to identify resonances in spectral ratios of the three component data of one station which were modeled as resonances due to P waves in the water layer between the ice and seafloor.

\begin{figure}
\includegraphics[width=25pc]{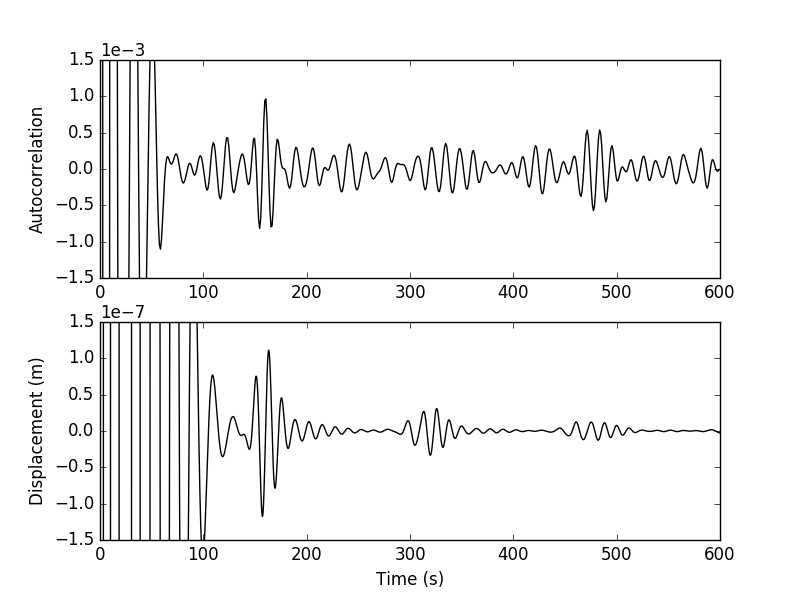}
\caption{Autocorrelation function (top) calculated from one week of simulated noise using the Model C seismicity and the Europa model with 20 km thick ice shell and low $Q$ compared to the zero offset displacement trace calculated using a single vertical force of $10^{10}$ N (bottom).  Both traces are filtered between 0.05 and 0.1 Hz.}
\label{ACFfig}
\end{figure}

In order to test the feasibility of such an approach with a noise source like the one modeled in this study, we calculated autocorrelation functions, following the approach of \citet{Huang+2015}. For this test we used the vertical component of a 1-week noise record in the 20-km ice shell model with low $Q$. The resulting response (fig.~\ref{ACFfig}, top line) shows a clear arrival near 175 seconds, which represents the reflection from the ocean floor.  There also appear to be arrivals that correlate with the first and second multiples of that reflection, which are clearly shown in the trace calculated using Instaseis for a co-located source and receiver (fig.~\ref{ACFfig}, bottom line), although the multiples cannot be easily distinguished from the background oscillations in the autocorrelation function.  Details of the autocorrelation function depend strongly on choices of filtering and autocorrelation window length, and while this particular model appears to show the multiples, we can generally only robustly see the first reflection using other noise records and Europa structure models. Even this first reflection, however, would allow for rapid determination of total ocean depth, even in the absence of any other tectonic events.  

Identification of such signals requires careful processing and filtering of the data, so it may be possible to find other signals in the horizontal components or using other processing that may also constrain other values of interest, like the ice shell thickness.  Resonances may also be easier to detect with spectral ratios, as shown by data from the Amery Ice Shelf \citep{Zhan+2013}.  In this test, though, we did not include any estimate of the observing instrument self-noise.  As illustrated by \citet{Zhan+2013}, however, coherent noise (due to propagating waves) needs to exceed incoherent noise across the filtered frequency band in order to obtain reliable autocorrelation information.  That will clearly be very challenging or impossible in the frequency band shown in figure~\ref{ACFfig} based on the mean PSD estimates discussed in this study, unless activity level is at or above the highest levels explored here and a very sensitive instrument, such as the STS2 or the VBB instrument from the InSight mission \citep{Lognonne+2015}, is used.  Realistically, such observations are likely only going to be possible if it will be possible to select data from high noise periods that may exist due to temporal variability due to diurnal tidal variations or other processes.

\subsection{Impact of $b$ values}
\label{bvalue}

While a $b$ value of 1 is justified based on terrestrial datasets, it is reasonable to consider whether Europa seismicity may have different behavior.  For example, different means of grouping lunar event catalogs have resulted in very different estimates of $b$ values for lunar catalogs.  \citet{Lammlein+1974} systematically identified events in Apollo data via waveform matching and separated them into likely moonquakes and impacts.  Both catalogs exhibited $b$ values greater than 1, with the impacts at a value of 1.3 and the moonquakes at 1.78.  A value that high has significant implications for the method employed in this study.  They argue that the higher $b$ values may be characteristic of tidal triggering, leading to a relatively large number of small events compared to large events.  The authors, though, calculated their $b$ value by comparing the logarithm of peak amplitude at the stations with number of events, rather than magnitude.  This could introduce some bias into their estimate, as magnitude accounts for distance as well.  Larger, distant events could then be grouped with smaller closer events, while smaller, distant events may be missed altogether.  Overall, it is not clear how large this bias would actually be or even what direction it would change our estimates.  Regardless, high $b$ values indicate larger relative amounts of small event relative to the largest events, which would reduce our expectation of recording large events, but increase the number of small, constantly ongoing events.
On the other hand, \citet{Nakamura1977} analyzed a subset of larger distant events, which he categorized as High Frequency Teleseismic (HFT) events.  For these events, he calculated ``lunar magnitude'', defined in a similar fashion as Earth magnitude scales, and calculated very low $b$ values of 0.5.  A catalog with such a low value of $b$ would lead to almost all energy being released in a few large events, and a comparatively small number of small events.

Fracturing events in ice may also have different statistical characteristics than those in rock.  As reviewed by \citet{Podolskiy+2016}, icequake catalogs on Earth show a wide range of $b$ values from less than 1 to greater than 2, even among studies looking at similar types of events.  Most studies, however, seem to cluster either around a $b$ value close to 1, or a higher value closer to 2.

\begin{figure}
\includegraphics[width=15pc]{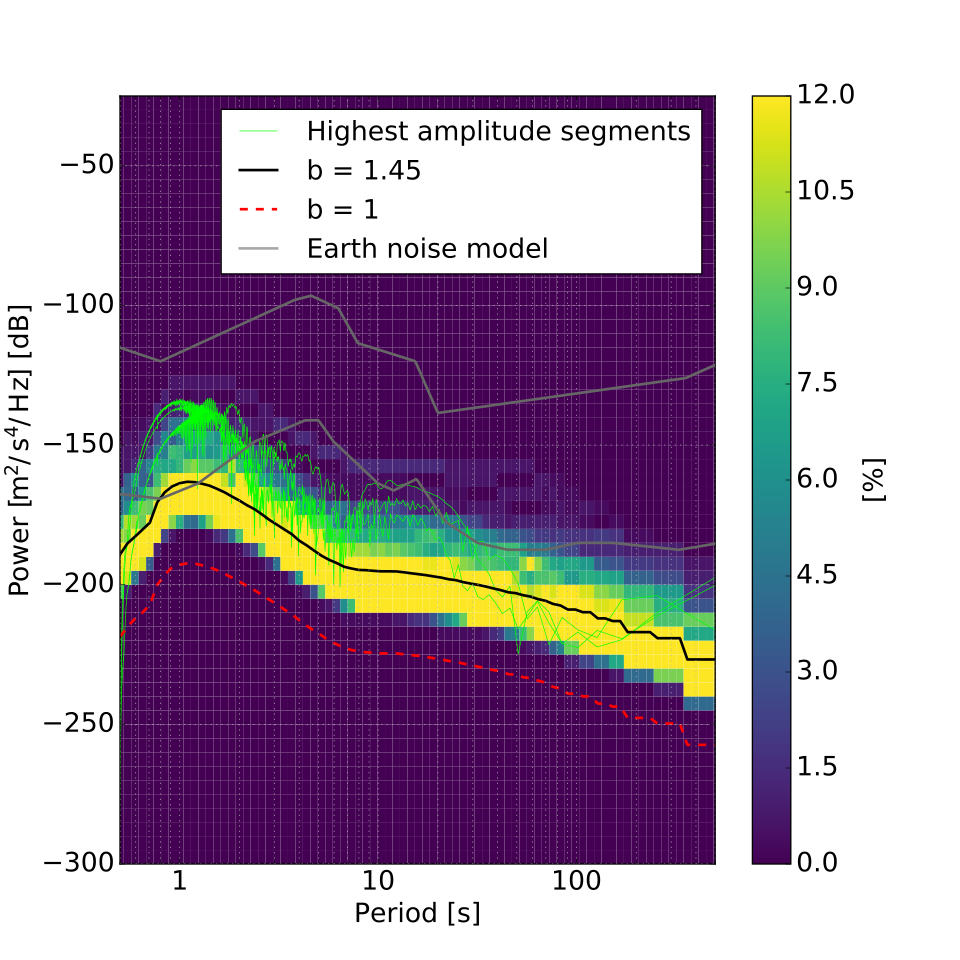}
\caption{Same as figure~\ref{ppsdfig}, but using a catalog with the preferred cumulative seismic moment and maximum event size, but with a $b$ value of 1.45.  The mean PSD for the case with b of 1 is plotted as a red dashed line.}
\label{highb_fig}
\end{figure}

The wide scatter in estimated $b$ values in lunar and ice settings suggest the difficulty in estimating such statistics using datasets that are comparatively limited in space and time compared to the Earth tectonic activity catalogs.  However, they do provide a strong suggestion that higher $b$ values may be appropriate in both tidally triggered settings as well as in ice.  While many resolved values in previous studies are between 1.5 and 2, we choose to initially explore the effect of increasing the value to 1.45 (fig.~\ref{highb_fig}).  For values greater than or equal to 1.5, equation~\ref{G92-2} is no longer valid.  This is because for such high $b$ values, energy release is no longer dominated by the largest events due to there being a decrease of greater than an order of magnitude in occurrence frequency for each order of magnitude increase in seismic moment.  In order to constrain the cumulative moment in this case, you actually need to determine the minimum event size, rather than the maximum.  Minimum event size, however, is not easy to define by any macroscopic observation.  For this reason, we choose to only test a value of $b$ less than 1.5.

As demonstrated in figure~\ref{highb_fig}, a $b$ value approaching 1.5 greatly increases the number of frequent small events, which raises the background noise level.  The mean PSD of the noise follows the same spectral characteristics as the previous estimates, but is approximately 20 dB higher at all frequencies.  The hours with the highest amplitude signals, however, remain at similar amplitude as in the original estimates, as we have not increased the number of large events.  This increases the chance of useful results from recordings of ambient noise, but possibly indicates lower signal to noise ratio for the largest events.

\subsection{Ocean noise}
\label{oceannoise}
On the Earth, ambient noise at most stations is dominated by microseisms, which originate in the ocean due to pressure variations at the ocean floor related to wave interactions \citep{Longuet-Higgins1950}.  While a subsurface ocean will not have the wind-driven gravity waves observed in the Earth's ocean, tidal deformation will generate motions in the ocean.  A study of the turbulent flow produced in the ocean suggests that there may be significant radial flow velocities approaching 2.5~m/s immediately below the ice shell \citep{Soderlund+2014}.  We convert these velocities to dynamic pressures acting on the base of the ice shell using the relationship $P=\frac{1}{2} \rho U^2$, where $P$ is pressure, $\rho$ is fluid density, and $U$ is radial flow velocity. Associated pressure variations are a few kPa, which is comparable to pressure variations at the floor of Earth's ocean.

\begin{figure}
\includegraphics[width=25pc]{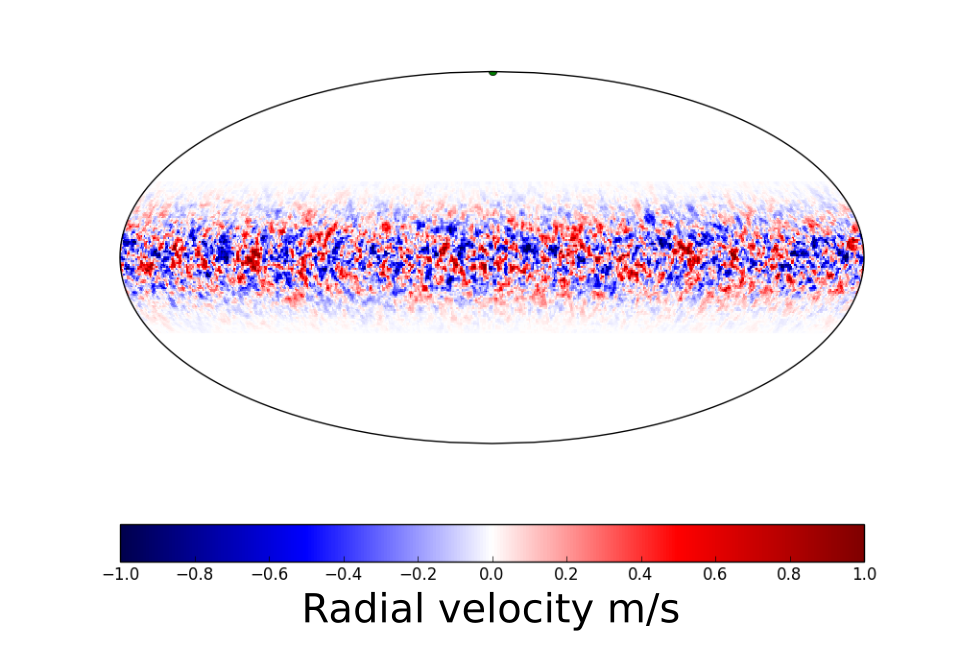}
\caption{An example random radial ocean velocity field time slice generated as described in the text with parameters chosen to match the general characteristics of the model of \citet{Soderlund+2014}.}
\label{oceannoisemap}
\end{figure}

The simulation of \citet{Soderlund+2014} only modeled motions with periods longer than $\sim$1000~s, so the excitation occurs at significantly longer periods than we have focused on with the ice tectonic sources.  To model this in the range of frequencies discussed in this study, we create a random radial velocity model (fig.~\ref{oceannoisemap}), which we convert to dynamic pressure at the base of the ice shell, as an input source to Instaseis.  This allows us to generate noise time series comparable to those we calculated for the ice tectonic sources.  We generate a random velocity field that is correlated in both space and time as defined by a von Karman autocorrelation function, as has long been used in defining randomly perturbed seismic media \citep[e.g.][]{Sato1982}, with correlation lengths of 50 km in latitudinal and longitudinal directions and 1000~s in time.  The structure is confined to within 20$\degree$ of the equator, compatible with the larger amplitudes observed near the equator in the model of \citet{Soderlund+2014}, as seen in the example time slice shown in figure~\ref{oceannoisemap}.  The correlation lengths in space and time chosen here produce a time-varying radial velocity model that is very similar in characteristics to the wavelengths and time variation in the model from \citet{Soderlund+2014}. The amplitudes of velocities are set to vary on the scale of $\pm$1 m/s in order to be comparable with the model of \citet{Soderlund+2014}.

\begin{figure}
\includegraphics[width=25pc]{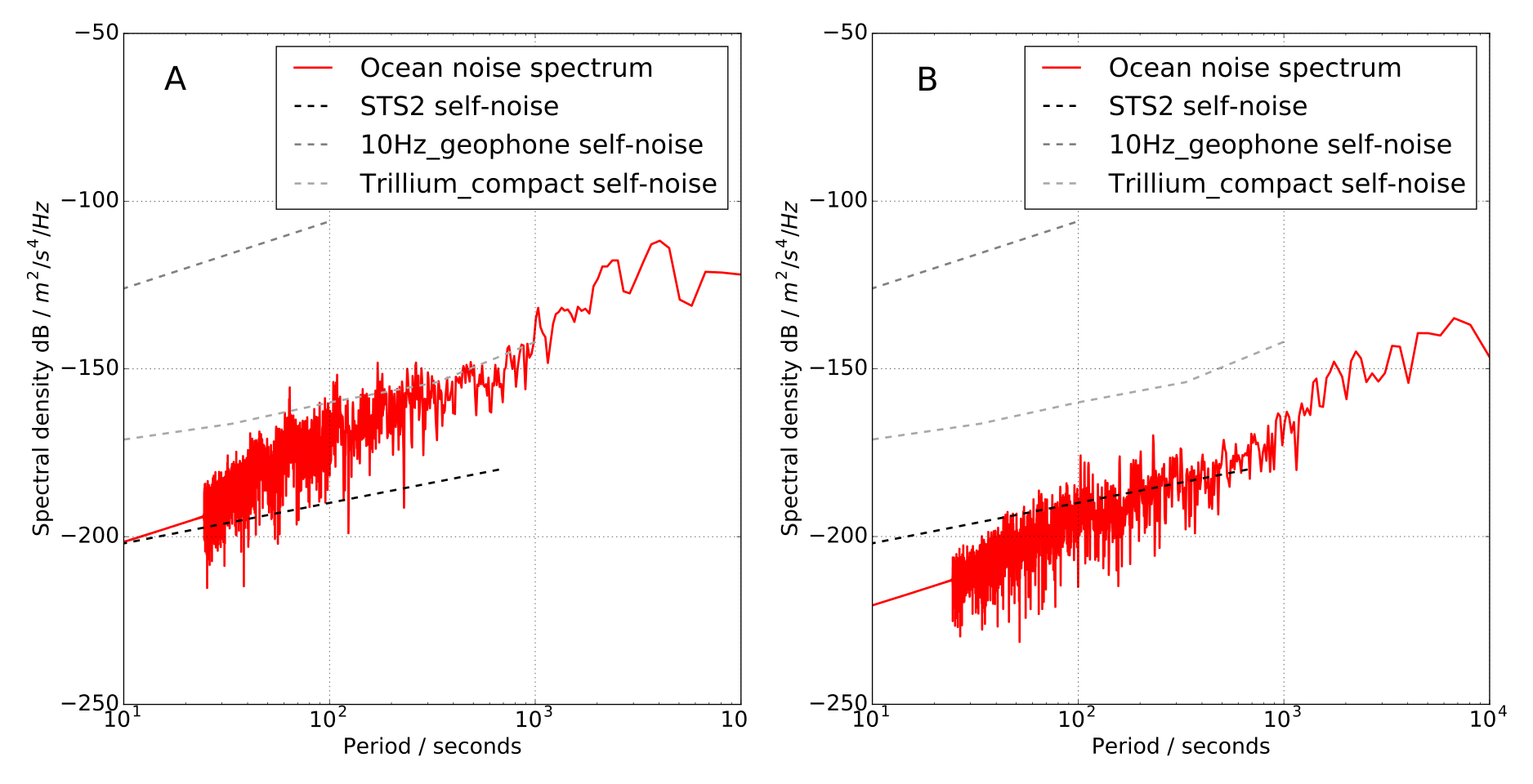}
\caption{Calculated noise spectra from modeled ocean pressure variations at the base of the ice shell in the 20 km model with high $Q$ at the equator (A) and north pole (B) compared with instrument self-noise curves.}
\label{oceannoisespectra}
\end{figure}

The resulting seismic noise spectra produced by such a model are shown for the 20~km thick model with high $Q$. The noise amplitude depends on the latitude of observation, with significantly larger amplitudes at the equator than the poles (fig.~\ref{oceannoisespectra}).  The equatorial signal is comparable with the noise floor of the SP or Trillium compact instruments.  In each case, the spectrum drops off at higher frequencies, and this drop-off is a function of the drop-off of the spectrum of the von Karman source-time functions.  If we used another method to extrapolate to higher frequencies, it is possible that this slope could change, but this indicates that ocean turbulence may create enough pressure signal to be observable with potential planetary seismic instruments.

As discussed by \citet{Zhu+2017}, however, the values of the ocean thermal diffusivity and vertical temperature gradient utilized by \citet{Soderlund+2014} may be unreasonably large.  If this is the case, turbulent flow velocities could be considerably lower than estimated here.  In fact, they may be low enough that global variations in ice shell thickness produce a stratified layer of lower salinity water in the region of ice melting \citep{Zhu+2017}.  Such a stratification would act to further dampen radial flow velocities at the base of the ice shell, decreasing the estimated ocean noise.

This model also only considers dynamic pressure forcing on the base of a smooth spherical shell, which would not generate significant excitation of Love waves and other SH modes.  However, on the Earth, the background long period seismic hum, which is believed to primarily originate in the oceans, shows excitation of Love waves and toroidal modes \citep[e.g.][]{Kurrle+2008,Nishida2013}, which may be caused by shear tractions associated with ocean floor topography \citep{Nishida2013}.  This implies that realistic topography of the base of the ice shell will be important for understanding noise from the ocean on Europa.

\subsection{Effects of scattering}
\label{scattering_sec}
All the modeling done to this point has assumed simple 1D models of structure for Europa.  Real data, however, are affected by small-scale structure which scatters seismic energy from the simple geometric paths predicted in a layered 1D model.  For example, our only other high-quality planetary dataset including clear tectonic events, the Apollo catalog of lunar seismic data, is dominated by scattering originating in a regolith layer that is highly fractured, but with very little intrinsic attenuation \citep[e.g.][]{Goins+1981}.  Such scattering can greatly change the character of a seismic record, reducing the amplitude of geometric phase arrivals as energy is scattered from the geometric path, while simultaneously producing extended codas (which on the Moon can continue for an hour or longer) representing energy that propagated longer distances due to off-path scattering.  In the absence of attenuation, such effects should not cause significant changes, however, to the frequency characteristics of noise, as the scattering simply shifts energy of a given frequency in time from geometric arrivals to extended codas.  Spectra estimated over sufficiently long durations should show similar characteristics in this case.  In the presence of attenuation, though, the longer scattered paths will allow for more energy loss due to intrinsic attenuation.

\begin{figure}
\includegraphics[width=15pc]{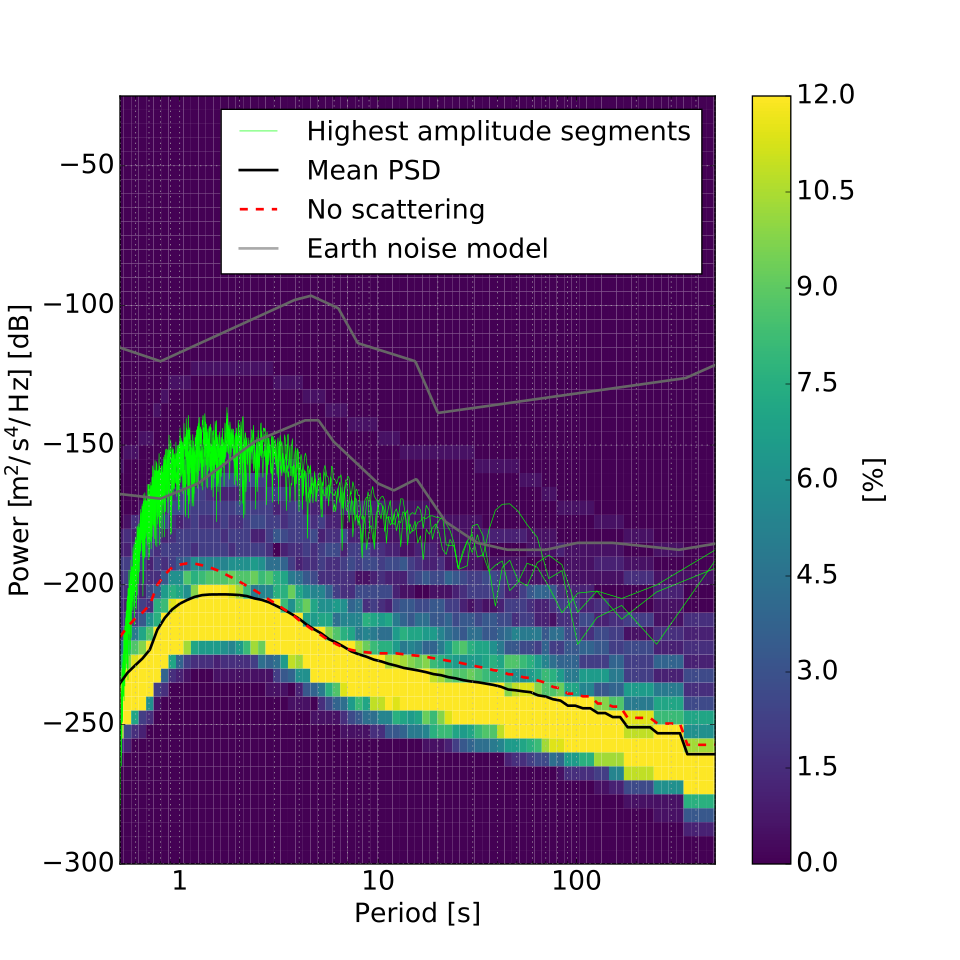}
\caption{Same as figure~\ref{ppsdfig}, but using a model including heterogeneity to simulate moderate scattering within the great-circle path between source and receiver.  The mean PSD for the case with no scattering is plotted as a red dashed line.}
\label{scattering_fig}
\end{figure}

Modeling strong 3D scattering like the Moon's requires either very finely sampled (and thus computationally intensive) numerical wave propagation or clever statistical techniques like a seismic phonon method \citep[e.g.][]{BlanchetteGuertin+2015}.  Such approaches can include multiply scattered energy from anywhere in the planet.  This modeling, however, is beyond the scope of this initial study of the Europa noise environment.  We can model simpler scattering due to relatively long-wavelength structure within the great-circle path between source and receiver using the AxiSEM/Instaseis approach of this study.  

The scattering model used here is discussed more thoroughly in \citet{Stahler+2017}. We implement a von Karman random medium with a correlation length of 5~km and velocity variations of 10\%.  As before, we calculate an hour-long waveform database using AxiSEM and then use Instaseis in combination with our seismicity catalogs to estimate a continuous noise record which is used to estimate a PSD for the noise (fig.~\ref{scattering_fig}).  We show the results for a scattering model superimposed over the high Q model with a 5 km ice shell.  The general character of the noise is similar, however there is a significant reduction in the mean PSD amplitude for periods shorter than a few seconds.  Some of this effect may be partially due to longer scattering paths pushing energy outside of the hour-long waveform records used for our noise calculation, but it is likely that this is mostly due to greater loss of energy to attenuation due to longer scattered path lengths at higher frequencies.  Large events still show a similar offset by 30--50 dB as seen in the unscattered records, but the peak energy is shifted to slightly lower frequency compared to the unscattered case.

\subsection{Effects of regolith}
On an airless body, a surface regolith that is highly fractured with void-filled cracks formed by impact gardening or tidal tectonic fracturing is likely, although the depth of such a layer on Europa is difficult to estimate.  Coherent backscatter of Earth-based radar measurements suggest a high porosity (25-75\%) layer extending to depths of up to a few meters at most \citep{Black+2001}, which is consistent with scatterers that are either filled with void or contrasting ice.  This could represent an impact gardened regolith, which can be modeled based on craters to extend up to a few meters deep \citep[e.g][]{Moore+2008}.  \citet{Eluskiewicz2004} proposed a regolith layer could reach a thickness in excess of 1 km based on estimates of compaction timescales as a function of depth.  A more recent response to that work has been updated to more accurately model development of regolith due to tidal fracturing (rather than impact gardening) along with modeled temperature profiles to determine where ice creep rates would be sufficient to close any open pores.  A large range of models was possible with depths of such a regolith varying between 0.5 and 3 km, with porosities varying between 1 and 22\% \citep{Aglyamov+2017}.  Whatever the thickness, such a layer would likely introduce more intense 3D scattering than that modeled in section~\ref{scattering_sec}.  Modeling such a layer is beyond the scope of the current study as it will require computationally intensive 3D numerical wave propagation codes or explorations using stochastic methods based on radiative transfer theory \citep[e.g.][]{GIllet+2016} or the seismic phonon method \citep[e.g.][]{BlanchetteGuertin+2015}, but we can consider qualitatively how such a layer may affect seismic data recovered from a landed Europa mission.  

Such a layer will likely act to reduce the amplitude of any Crary waves observed in the data, since the Crary waveguide relies on homogenous properties of the ice shell giving rise to perfect reflections at the surface and base of the ice shell for SV waves with a horizontal slowness equal to that of a P wave propagating in the ice shell \citep{Crary1954,Vance+2017a,Stahler+2017}.  However, the scattering and coda such a layer produces could be used to increase other kinds of science return from seismic data.  For example, the scattered seismic energy either in ambient noise or coda of phase arrivals could be used to extract high frequency Rayleigh wave ellipticity information (greater than 1 Hz), which has been proposed for use in constraining near-surface structure on Mars using data from the upcoming InSight mission \citep{KnapmeyerEndrun+2016}.  Further investigation of possible effects of regolith scattering will likely be essential in order to understand any returned seismic data from Europa.

\subsection{Spatiotemporal variation}
For simplicity, we have assumed that seismicity follows a statistical Gutenberg-Richter relationship that is stationary in both time and space.  The tidal stresses on Europa, though, vary as a function of time and space during each orbital cycle around Jupiter \citep[e.g.][]{Greenberg+1998}.  This implies that the noise estimated in this study can only be considered a mean value, and the actual levels will vary depending on the choice of landing site and within each $\sim$85 hour tidal cycle.  Consideration of this tidal variation as well as likely spatial variation in ocean noise generation as discussed in section~\ref{oceannoise} will be critical in landing site selection to maximize seismic data return.

\subsection{Instrument requirements}
Following \citet{Lee+2003}, the recently released report of the Europa Lander Science Definition Team \citep{Hand+2017} argued for a noise floor of -35 dB with respect to a velocity of 1 $\mu$m/s in order to establish a preliminary instrument requirement.  If we treat this floor as flat in velocity, and convert to the acceleration power spectral density relative to 1 m/s$^2$ used in this study, this would correspond to approximately -175 dB at 1 Hz and -185 dB at a period of 10 s, which is comparable to the mean PSD values for the ``model C'' seismicity model in the high $Q$ Europa models.  This suggests our approach is broadly consistent with previous noise estimates, and further suggests that both estimates of the noise floor indicate that a high frequency geophone is likely not sufficient to meet science requirements for a landed Europa mission, without relying on future modeling and observations to allow us to specifically pick a landing site near expected activity.  Meanwhile an instrument with sensitivity similar to the Trillium Compact or the InSight SP instrument will more likely meet science requirements based on the homogenous modeling developed here.  Specifically, it should be able to reliably record important phase signals from larger events (i.e. signal), and may be able to record the background ambient noise if the actual seismic activity is in the higher range of our estimates, or in time periods of higher activity in the tidal cycle.

This study primarily focused on the noise recorded on the vertical component, which often has the lowest noise due to local site effects in Earth settings, but future work will also need to focus on noise and signals from the horizontal components in order to more fully evaluate the relative utility of sending a 3 component instrument or simply a 1-axis vertical instrument.  This will also help inform mission design on requirements of alignment of sensors, such as the need for a leveling system or control of horizontal component azimuthal orientation.  Polarization information is essential for determination of back azimuth in single station location techniques \citep[e.g.][]{Panning+2015}, and access to wave types with horizontal polarization provides important constraints on relevant parameters \citep[further discussion in][]{Stahler+2017}, and so a 3 component instrument should have significant advantages.  Further work, though, will be required to look not just at ambient noise sources, but also details of installation and lander noise.

\section{Conclusions}
In order to estimate the likely seismic activity and noise levels for an instrument on Europa's surface, we explore a range of seismicity models that follow a Gutenberg-Richter relationship.  The seismic activity level in such models depend on the cumulative seismic moment release and maximum event size.  Given a range of reasonable values for these parameters scaled from observed activity levels on the Moon, we generate catalogs, and then use them to generate models of seismic activity and noise using numerical wave propagation codes through thermodynamically consistent models of Europa's interior structure.

Given this range of models, we show that most reasonable models show background noise levels well below the sensitivity of a high frequency geophone, but potentially measurable by more sensitive instruments particularly for the higher seismicity models.  The amplitudes of the largest events observable in a given period of a few weeks are likely observable by more sensitive broadband instruments analogous to a Trillium Compact or the InSight SP instrument.

We demonstrate the potential of auto-correlation of such noise records to determine the ocean depth.  We also explore the possible amplitude of noise generated by turbulent flow in the subsurface ocean due to tidal motions.  Such a noise source may be observable with reasonable planetary seismic instruments at longer periods.

\acknowledgments
\begin{sloppypar}
The authors acknowledge computational support in the project pr63qo "3D wave propagation and rupture: forward and inverse problem" at \textit{Leibniz-Rechenzentrum} Garching. SCS was supported by grant SI1538/4-1 of Deutsche Forschungsgemeinschaft \textit{DFG}.  This work was partially supported by strategic research and technology funds from the Jet Propulsion Laboratory, Caltech, and by the Icy Worlds node of NASA's Astrobiology Institute (13-13NAI7\_2-0024).  Noise waveform records and seismic catalogs are available via GitHub at http://github.com/mpanning/EuropaNoise.  Axisem waveform databases are maintained by SCS and can be accessed for use in Instaseis scripts via http://instaseis.ethz.ch/icy\_ocean\_worlds/.  Sound files in supporting information were created with Matlab scripts from Zhigang Peng (http://geophysics.eas.gatech.edu/people/zpeng/EQ\_Music/).   Work by MP was started at the University of Florida, and was completed at the Jet Propulsion Laboratory, California Institute of Technology, under a contract with the National Aeronautics and Space Administration. Reference herein to any specific commercial product, process, or service by trade name, trademark, manufacturer, or otherwise, does not constitute or imply its endorsement by the United States Government or the Jet Propulsion Laboratory, California Institute of Technology.
\end{sloppypar}

\listofchanges

\end{document}